\documentclass[grl]{AGUarxiv}

\usepackage{lineno}

\usepackage{graphicx}
\authorrunninghead{LEMMEN AND KHAN}

\titlerunninghead{SIMULATION OF INDUS VALLEY NEOLITHIC}

\authoraddr{C. Lemmen,
Institute of Coastal Research, 
Helmholtz-Zentrum Geesthacht,
Max-Planck-Str.~1, 21205~Geesthacht, Germany,
(carsten.lemmen@hzg.de)}

\authoraddr{A. Khan,
Institute of Coastal Research, 
Helmholtz-Zentrum Geesthacht,
Max-Planck-Str.~1, 21205~Geesthacht, Germany,
(aurangzeb.khan@hzg.de)}

\begin{document}

\title{A simulation of the Neolithic transition in the Indus valley}

\authors{Carsten Lemmen and Aurangzeb Khan}

\begin{abstract}
The Indus Valley Civilization (IVC) was one of the first great civilizations in prehistory.  This bronze age civilization flourished from the end of the fourth millennium~BC.  It disintegrated during the second millennium~BC; despite much research effort, this decline is not well understood. 
Less research has been devoted to the emergence of the IVC, which shows continuous cultural precursors since at least the seventh millennium~BC.  To understand the decline, we believe it is necessary to investigate the rise of the IVC, i.e., the establishment of agriculture and livestock, dense populations and technological developments 7000--3000~BC. 
Although much archaeological information is available, our capability to investigate the system is hindered by poorly resolved chronology, and by a lack of field work in the intermediate areas between the Indus valley and Mesopotamia.  We thus employ a complementary numerical simulation to develop a consistent picture of technology, agropastoralism and population developments in the IVC domain.   Results from this Global Land Use and technological Evolution Simulator show that there is (1)~fair agreement between the simulated timing of the agricultural transition and radiocarbon dates from early agricultural sites, but the transition is simulated first in India then Pakistan;  (2)~an independent agropastoralism developing on the Indian subcontinent; and (3)~a positive relationship between archeological artifact richness and simulated population density which remains to be quantified. 
\end{abstract}

\begin{article}
\section{Introduction}

The Indus Valley Civilization (IVC), often termed Harappan civilization after its major type site, flourished  along the banks of the river Indus and its tributaries, including the adjacent coastal areas, between the fourth and second millennium~BC.  The IVC is characterized by urban centers, bronze technology and seals,  trade networks with Mesopotamia and Arabia, and an as yet undeciphered writing system \citep{Shaffer1992,Possehl1998,Kenoyer2008}.  Much research has been devoted to the disintegration of the IVC during the second millennium~BC, and much less to the emergence of this great civilization which shows continuous cultural precursors at least since the seventh millennium~BC, such that some authors prefer to speak of the Indus valley cultural tradition \citep[e.g.,][]{Kenoyer2006}.

\begin{figure*}
\centering\noindent\includegraphics[width=\hsize]{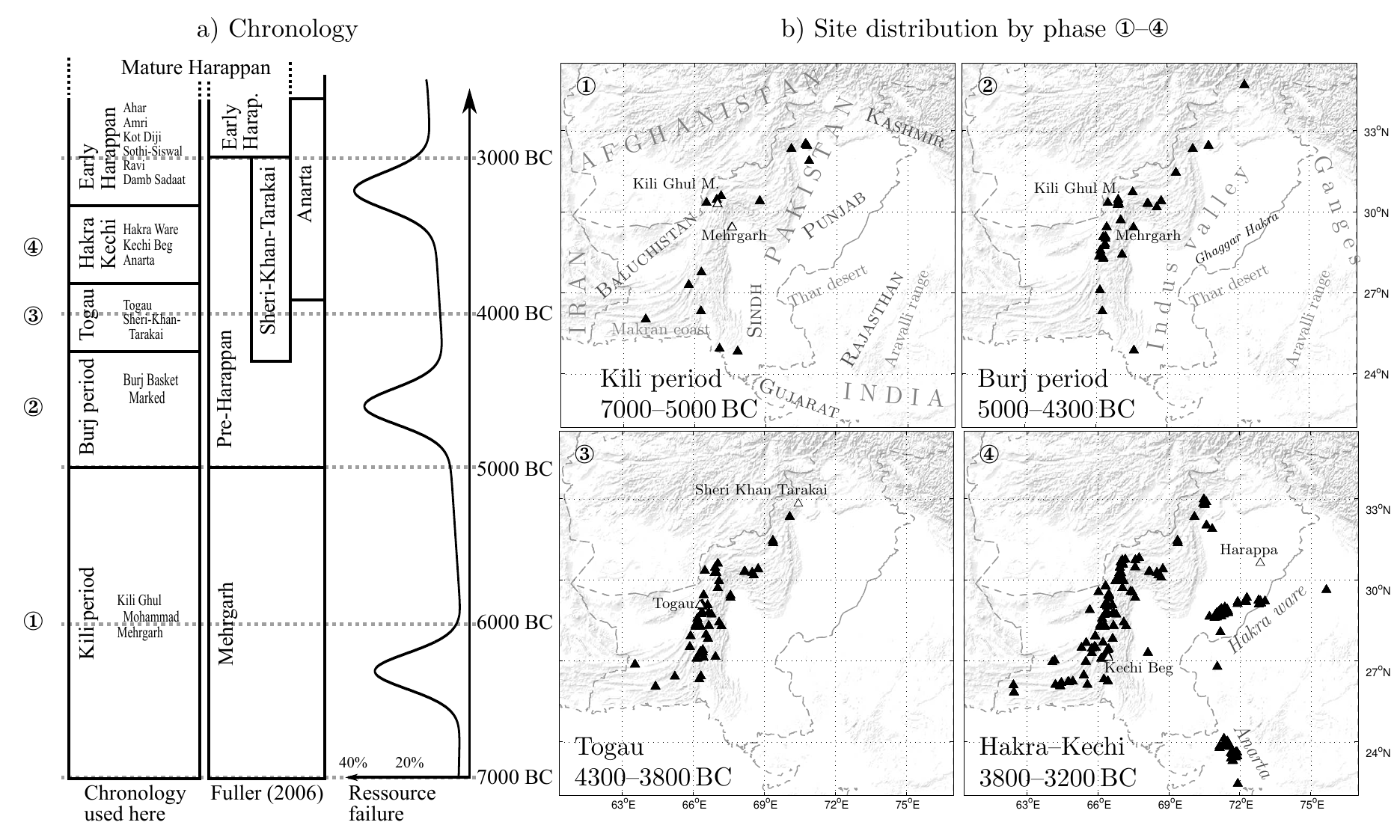}
\caption{Temporal and spatial domain for this study:  a)~Neolithic chronology of Baluchistan, the Indus valley, and Gujarat based on \citet{Possehl2002} with updates from \citet{Fuller2006}.  Relative resource shortage is based on palaeoproxy evidence from \citet{Lemmen2012}  and a transient Holocene climate simulation by \citet{Brovkin2002}. b-e)~Geography, topography and site distribution of artifacts \citep{Law2008} typologically associated with the Neolithic.}
\label{fig:chrono}
\end{figure*}

The Indus valley is one of the two great river basins on the Indian subcontinent, separated from the Ganges valley to the east by the Aravalli mountain range and the Thar desert (Figure~\ref{fig:chrono}b).  The region has been occupied by anatomically modern humans for at least 34,000, but possibly $>$80,000~years \citep{Petraglia2010}.  The late Paleolithic and the Mesolithic are visible in stone and blade industries; microlith blades indicate composite tools, differentiated occupation levels and a broad-spectrum diet of foraging people from 26,000 years before present (a\,BP) until at least the second millennium~BC, well into the Bronze age \citep{Petraglia2009}.

Neolithic subsistence is based on a combination of barley and wheat dominated agriculture and herding of sheep, goats, and cattle.  Domesticates are of mixed origin: barley and wheat have no local wild progenitor, but zebu cattle bears a clear signature of local domestication \citep{Chen2010,Fuller2011}.  The beginning of the Neolithic on the Indian subcontinent is regionally diverse and ranges from 6500~BC in Baluchistan \citep{Jarrige1995} to 3000~BC  or later on the central Indian plateau and in the Himalaya mountain range \citep{Samuel2000,Boivin2008}. Our focus here is on the period which spans from the Neolithic to the beginning of the Bronze age of the IVC (7000--3200~BC), i.e., on the early food production and regionalization eras, or the Mehrgarh phase and parts of the early Harappan phase according to the different chronologies \citep{Sharma1980,Fuller2006,Kenoyer2008}. 

\citet{Possehl2002} separated these phases and eras of the pre-Harappan Neolithic into two stages (Figure~\ref{fig:chrono}a): Early Neolithic and Developed Neolithic.  The Early Neolithic is divided in the Kili Ghul Mohammad (KGM) and the Burj Basket-Marked (BBM) periods. Sites attributed to these periods are found in the western borderlands of the greater Indus region and cluster in the hills and piedmont of Baluchistan (Figure~\ref{fig:chrono}\,b1--b2).  During the BBM, soft pottery and handmade pottery are introduced.  Deposits of trash with burnt pebbles, ash, animal bones, bone tools, hammer stones, polishers colored with red ochre, and a large collection of blades, cores, and flint debris have been discovered, and point to leather making, basket making, or weaving.  Domesticated wheat, barley and cotton were used, pulses may have been present; the use of copper is visible in the form of arrowheads and beads \citep{Moulherat2002,Fuller2011}.

The Developed Neolithic is characterized by improved pottery, developed agropastoral communities, growth, continuity, and geographical expansion.   
During its first period (here defined as both Togau and Sheri Khan Tarakai, SKT, Figure~\ref{fig:chrono}\,b3), the settlement patterns show growth of village life; technological innovations include the use of gold, the manufacture of compartmented seals, glazed steatite, and beads. An eastward expansion is visible in sites emerging along the ancient Ghaggar Hakra river.  The second period of the Developed Neolithic (Figure~\ref{fig:chrono}\,b4) comprises several cultural complexes.  In the Kechi Beg complex in Baluchistan wheat replaces barley; an irrigation canal near Mehrgarh may date to this period \citep{Kenoyer2008}.  On the Punjab plain, a change in the assemblage of pottery types characterizes the Hakra Ware complex: Microlithic tools are  abundant, and sites represented by a light scatter of pottery without a buildup of a midden point to seasonal occupations \citep{Possehl2002}.  At the same time, the Anarta complex appears in Gujarat with distinct pottery styles. 

Our capability to investigate the system more closely and consistently in the regional context is hindered by the poorly resolved chronology, and by a lack of field work in the intermediate areas between the Indus valley and Mesopotamia.  We thus employ a complementary numerical simulation to the IVC domain. \citet{Ackland2007} simulated the Neolithic transition in the IVC with an advancing frontier spreading from Southwest Asia, and a competition between migrating original farmers and resident converts.  After 5000 simulation years (9000--4000\,BC), their simulated spread of agriculture into the Indian subcontinent exhibits a demarcation line along the Indus river, separating original farmers to the west from converts to the east. 

A local South Asian agricultural center was considered in the numerical model by \citet{Patterson2010}.  They assumed a background hunting-gathering population, and initial farmer populations near Mehrgarh and at the eastern edge of their model domain. After 2000~simulation years, the population in the Indus valley is dominated by expanding original farmers, a lesser share of converts, and a tiny fraction of remaining hunter-gatherers.

We go beyond the biophysical approaches by \citet{Ackland2007} or \citet{Patterson2010};  in our Global Land Use and Technological Evolution Simulator \citep[GLUES, e.g.,][]{Lemmen2011f}, we include socio-technological innovation in addition to migration, population growth and subsistence change to develop a consistent picture of the dynamics of technology, agropastoralism and population developments during the Mesolithic--Neolithic transition.  This simulated data is subsequently compared to a representative archaeological data set \citep{Law2008}, where the chronology of Neolithic sites is used to detect the regional transition from foraging to farming, and where the site occurrence frequency is used to detect the Neolithic demographic transition.

\section{Data and model description}

\subsection{Global Land Use and technological Evolution Simulator}
GLUES is a socio-technological model which hindcasts technological evolution, potential population density, and the timing of the  transition to agropastoralism for 685~world regions based on the geoenvironmental and cultural contexts, and innovation and adaptation of regional populations.  Each local population in an approximately country-size region utilizes its natural resources and interacts with its geographical neighbors via trade and migration.  The full model is described by \citet{Wirtz2003gdm}, with refinements of geography \citep{Lemmen2009}, migration and knowledge loss \citep{Lemmen2010}, and climate events \citep{Lemmen2012}; it has been validated against archaeological data by \citet{Lemmen2011f}.  Key concepts are summarized below:

Local societies are defined by their population density ($P$) and by three characteristic traits: (1) technology ($T$), which describes efficiency of food procurement; (2) share of agropastoral activities ($Q$) as the allocation pattern of manpower to farming and herding as opposed to foraging; and (3) economic diversity ($N$), a technology-related trait which represents the different subsistence styles of a community.  The temporal change of traits follows the direction of increased benefit for success (i.e.\ growth) of its associated population with a speed related to the trait's variance \citep{Fisher1930,Kisdi2010}.    
Population growth rate ($r$) depends on resource availability ($E$), subsistence intensity, overexploitation of resources, administrative overhead and health improvements; the latter four terms are functions of $P,T,N$, and $Q$, while the resource availability $E$ is provided externally.  $E$ is estimated from the net primary productivity, which is in turn based on \nobreak{Climber-2} \citep{Claussen1999} transient temperature and precipitation anomalies from the IIASA climatological data base \citep[International Institute for Applied Systems Analysis, ][]{Leemans1991}.    

Abrupt climate deteriorations modulate $E$; they are incorporated by using the extreme event anomalies found in 124~globally distributed palaeoclimate proxies collected and analyzed by \citet{Wirtz2010}; the estimation of resource failure from proxy climate is described by \citet{Lemmen2012}.   Figure~\ref{fig:chrono}a shows the average relative decline of $E$  in the Indus valley region; these anomalies are derived from regional abrupt changes found in oxygen isotope anomalies from  Kyrgyzstan \citep{Ricketts2001}, Tibet \citep{Fontes1996}, the Arabian Sea \citep{DooseRolinski2001,Gupta2003}, and Oman \citep{Fleitmann2007}.  

We set up the eight global model parameters and initial values such that  the simulation is able to hindcast an accurate timing and location of the early farming centers in Southwest Asia, northern China, and Mesoamerica \citep{Smith1998}, and a reasonable global pattern of the subsequent Neolithic transition. The simulation is started at 9500\,sim\,BC (simulation years~BC; equivalent to calendar years~BC, but emphasizes the artificial model time scale,  \citealt{Lemmen2011f}). All of the 685~biogeographically defined regions, including 32 around the Indus valley, are initially set with farming activity at $4\%$, $1/4$ established agropastoral economies and unit technology; this setup represents a low density Mesolithic population relying on a broad spectrum foraging lifestyle with low unintentional farming activity.   The regional simulated transition to agropastoralism is recorded when $Q$ exceeds $0.5$.

\subsection{Artifact database}
We used the Indus Google Earth Gazetteer (version August~2008) compiled by \citet{Law2008} for the geolocation of artifacts relating to the Indus valley.  From this database of 21687~metal and stone artifacts, we use here the cultural attribution and location of those 2028~artifacts and 368~sites that are in the spatial and temporal domain of our study.  For KGM, 18~sites contribute 90~artifacts to the data base; for BBM, there are 159~artifacts at 32~sites.   We combined the sites attributed to the Togau period with sites from the SKT (both with starting date 4300~BC, \citealt{Fuller2006}) and obtain 317~artifacts from 65~sites.  

The subsequent period is represented by the Hakra Ware, Kechi Beg, and Anarta complexes.  Combined, there are 1250~artifacts (396, 670, and 184, respectively) from 253~sites (99, 112, and 42, respectively).  While the main chronology is based on \citet{Possehl2002}, sites from the Anarta complex were included in this period based on \citet{Fuller2006}.  To our knowledge, this gazetteer is the most representative data set of lithic and metal artifacts of the Indus valley tradition.  It has been used to investigate the trade and distribution networks of the IVC by its author \citep{Law2011}.


In archaeological or palaeoenvironmental archives,  the decrease of the number of dates further back in time that is caused by gradual destruction of the archive by post-depositional processes results in a taphonomic bias  \citep{Surovell2009}. 
We applied to the artifact density data a bias correction derived from the comparison of volcanic eruption layers in the GISP2 (Greenland, no taphonomic bias) ice core with radiocarbon dated ash deposits; the correction factor $c$ depends on time $t$ (in a\,BP): $c(t)=5.73 \cdot 10^{6} (2175.4+t)^{-1.39}$  \citep{Surovell2009}.  This method was successfully validated with Canadian coin mintage data compared to a coin collection and applied to North American radiocarbon dates by  \citet{Peros2010}.


\section{Results and Discussion}
\begin{figure*}
\noindent\includegraphics[width=\hsize]{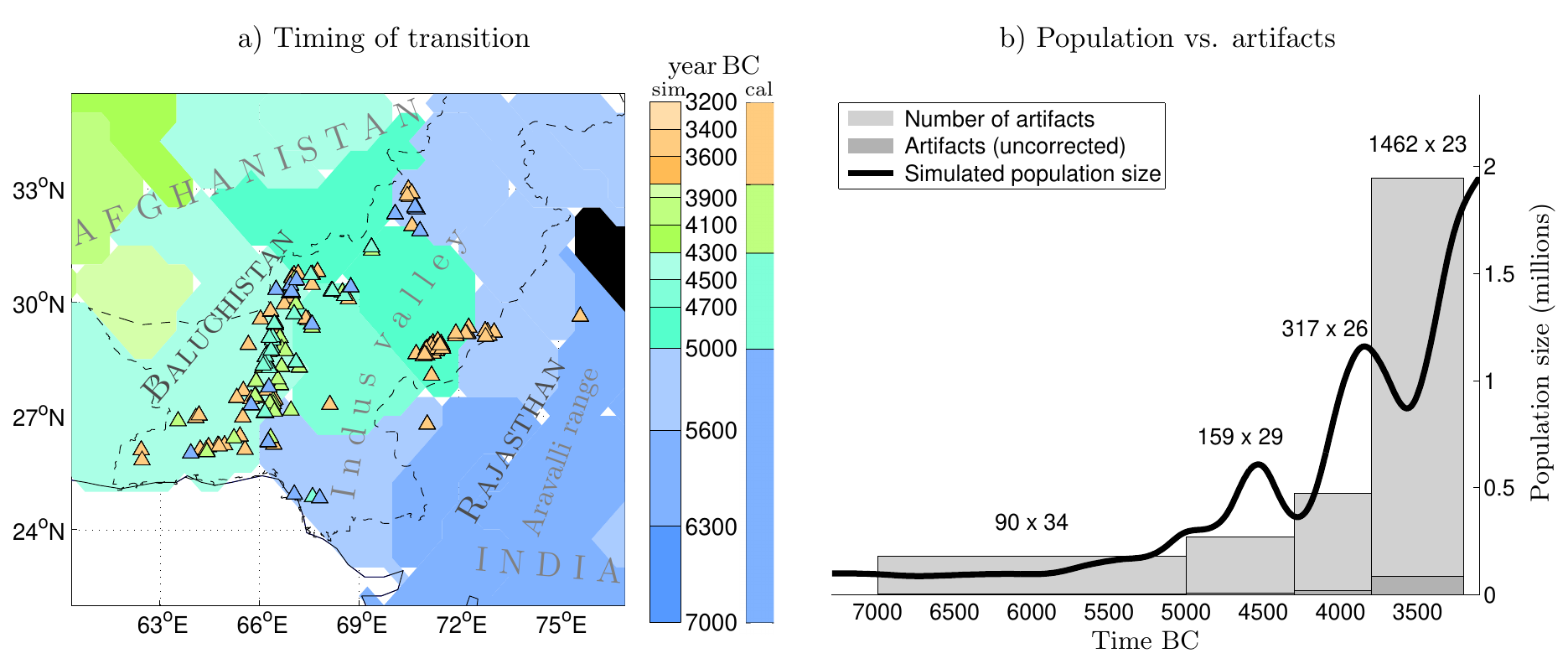}
\caption{a) Timing of the Mesolithic--Neolithic transition from a GLUES simulation (background color, simulation years~BC) and inferred from the presence of Neolithic sites (foreground triangles, calendar years~BC) in the greater Indus valley. b) Recovered artifact number in relation to simulated population size.  Histogram of artifact density in four pre-Harappan periods of the Indus valley and Baluchistan based on \citeauthor{Law2008}'s (\citeyear{Law2008}) Indus Google Earth Gazetteer, and corrected for taphonomic bias (bars, annotated with $n$ x $c$, where $n$ is the raw number of artifacts and $c$ the taphonomic correction factor);  GLUES-simulated population size in all areas within 200\,km from where Neolithic pre-Harappan artifacts were discovered (solid line).}
\label{fig:timing}
\end{figure*}
%
The simulated transition to predominantly agropastoralism-based subsistence occurs in the spatial domain of the Indus valley and surrounding areas between 6300 and 3800~sim\,BC (Figure~\ref{fig:timing}a).  There is a general east-west trend in the simulated transition dates.  Contemporary with the KGM, the earliest transitions are simulated in India (before 5600~sim\,BC), including Rajasthan, Gujarat, and the Ganges valley, and early transitions (before 5000~sim\,BC) in southern Sindh and northern Pakistan, including parts of the Punjab and Kashmir.   During this first period, 15 out of 32 simulation regions undergo the transition to agropastoralism.  Most of the remaining regions (Pakistan, eastern Afghanistan, eastern Iran and Makran coast)  transition in the BBM period before 4300~sim\,BC. Of these, agropastoralism is simulated first in the Punjab and northern Baluchistan (before 4700~sim\,BC), then in northern Sindh (before 4500~sim\,BC).  The simulated transition occurs latest, during the Togau period, in the eastern model domain, from central Afghanistan to the Pakistan-Afghanistan-Iran border triangle (4300--3800~sim\,BC).

A detailed and quantitative comparison is limited foremost by the inherent timing uncertainties in the typologically dated artifacts.  This is closely related to the secondary comparison difficulty, which is created by the lack of temporal continuity of samples from the Mesolithic to the Neolithic.  While we have shown how these issues can be addressed in the context of plenty and precise dates for Europe \citep{Lemmen2011f}, the comparison with dates from the Indian subcontinent can only be qualitative at this moment (see also \citealt{Patterson2010}, who make a qualitative comparison not with single sites,  but with approximate timing contours).  

In addition, the coexistence of Mesolithic and Neolithic subsistence in the archaeological record complicates the comparison between model and data, because the degree of both subsistence styles cannot be quantified with the current data.  This coexistence is evident in the long-term continued use of (Mesolithic) microlithic tools, transhumance life style, i.e., the seasonal back-and-forth movement of herders with their livestock, and the continuation of foraging practice well into the Neolithic \citep{Mughal1990,Fuller2011}, it is not resolved in the model, where agropastoral life style is usually dominant within less than 500~years after its emergence \citep{Lemmen2012}.

The model simulates first agropastoralism in northern Baluchistan and southern Sindh in the KGM period consistent with earliest sites in these areas. For the major part of Baluchistan, including the Mehrgarh site, the model places the transition in the BBM period, which is later than the earliest site dates, but agrees with dates from many other locations in Baluchistan.   The most obvious mismatch is the early simulated (KGM and BBM) transition along the Indus, the Ghaggar-Hakra and Punjab rivers, where site dates are absent until the Kechi Beg and Hakra periods.

While evidence for early archaeological sites may have been lost in the geomorphologically dynamic Indus river floodplain, the lack of early sites in Rajasthan is striking, considering that the model sees a biogeographically favorable environment and cultural setting for an early Neolithic: one should ask whether an agricultural center on the Indo-Gangetic divide is undetected so far.  First evidence for northern Indian agriculture could come from further east, from Lahuredawa in the middle Ganges plains, where evidence for domesticated rice appears as early as 6500~BC \citep{Tewari2008} (but see \citealt{Fuller2011} for a debate on the rice morphology).
Less favored by the model are the valleys along the Indo-Iranian plateau, where broad subsistence possibilities are seen as one precondition for the rise of the IVC and where agropastoralism arose before 6500~BC \citep{Jarrige1995}.  The model might underestimate the potential for agropastoralism in this area because of its coarse spatial scale.

From a regional perspective, the simulated east-west direction of agropastoralism contradicts the eastward trend seen in the site data. 
The westward direction had been suggested in earlier literature on the Indus valley tradition:  \citet{Wheeler1959} wrote of ``movement of the Neolithic from Burma if not behind.''  In their simulation of the spread of agriculture into the Indian continent, \citet{Patterson2010} assume such a Southeast Asian center in addition the Mehrgarh location.  While \citeauthor{Wheeler1959}'s (1959) earliest dates (1000~BC) have been pushed further back by more recent datings,  this westward view cannot be upheld from either artifacts \citep{Law2008} or the description of sites in central India (3000~BC, \citealt{Boivin2008}), the Indus plain (4000~BC, \citealt{Mughal1990}), northern India (5000~BC, \citealt{Sharma1980}), or Baluchistan (6500~BC, \citealt{Jarrige1995}); all of these demonstrate the opposite east- and then southward trend.

The simulated earliest agropastoralism in the Indus valley region is independent of the two prescribed early centers of agropastoralism in Eurasia, i.e., China and Southwest Asia; towards the end of KGM, however, a continuous band of early agropastoralism ranges from the northern Indian subcontinent through northern Southeast Asia to China (not shown).  This South and East Asian agricultural area connects to the Southwest Asian center during the BBM period.  The simulated independent emergence of South Asian agropastoralism corresponds well to evidence for independent domestication of cattle and rice \citep{Chen2010,Fuller2011b}.  Though not completely detached from Southwest Asian domesticates assemblage in the data, this separation is visible in the delayed simulated transition to agropastoralism in the area between the Indo-Gangetic and the Southwest Asian founder centers. 

Building on the earlier idea of \citet{Sanders1979}, we relate the number of artifacts in the database (corrected for taphonomic bias) to population estimates from our simulation (Figure~\ref{fig:timing}b).  We summed the total simulated population in each half-degree model grid cell within 200\,km distance to the find sites and obtain a population trajectory which shows an increase from $0.2$~million at 6000 sim\,BC to 2.5~million at 3500 sim\,BC; this corresponds to a population density of $0.1$ and $1.3$\,km$^{-2}$, respectively.  At 3500~sim\,BC  our estimate is higher than the one by \citet[][$0.2$\,km$^{-2}$]{Patterson2010}, as should be expected from our calculation from the surroundings (within 200\,km) of mostly settlement sites.  To first approximation, the increases in simulated population size and in reconstructed artifact density correspond: the model correctly represents the emergence of dense settlements.  Statistically, however, no meaningful relationship should be derived from only four different periods.

In summary, there is fair qualitative agreement between the reconstructed and the simulated transition to agropastoralism, both in the spatio-temporal distribution of agropastoral sites and simulation regions, as well as the overall artifact and population density.  Important lessons from the model are that it corroborates an independent agropastoral center, and that it suggests potential for earlier Indian agropastoralism. It will be left to further studies to quantify the agreement and to elaborate on the reasons, both in archaeology and in the simulation, of discrepancies.

\section{Conclusion}
We presented a novel numerical simulation of the pre-Harappan Neolithic of the Indus valley tradition in the context of representative radiocarbon dates from material culture.  Simulated population size is qualitatively reflected in artifact frequency in four pre-Harappan periods.  Within the uncertainties of the coarse chronology, the model predicts the spatio-temporal structure of the Neolithic transition of this area fairly well.  Our simulation points to a possible earlier Neolithic in northern India than seen in the data, and it corroborates an independent South Asian Neolithic.   For quantitative model--data comparisons and clarification of the connection to Southwest Asia, better chronologic control of the pre-Harappan material is needed as well as more evidence from the intervening Iranian plateau.

\section*{Supplementary material}
The simulated data have been permanently archived on and are freely accessible from PANGAEA (Data Publisher for Earth \& Environmental Science, \nobreak{http://www.pangaea.de}) as a netCDF dataset with reference 
``Lemmen, C.\ and Khan, A.\ (2012): Simulated transition to agropastoralism in the Indus valley 7500--3000\,BC. Dataset~\#779706, \textbf{doi:XXXX}''.  GLUES is open source software and can be downloaded from \nobreak{http://glues.sourceforge.net}.

\begin{acknowledgments}
We thank R.W.~Williams for publicly providing his extensive data set for further analysis; we thank three anonymous reviewers and the editor for their critical and helpful comments on earlier versions of this manuscript.  C.L.~acknowledges financial support from the DFG priority program Interdynamik (SPP\,1266), A.K.~acknowledges financial support from the Higher Education Commission of Pakistan and the PACES program of the Helmholtz Gemeinschaft.  This paper benefitted greatly from ideas presented and discussed at the AGU Chapman Conference on Climates, Past Landscapes, and Civilizations, held in Santa Fe, USA, March~2011. Maps were prepared with Natural Earth free vector and raster map data available from \nobreak{http://naturalearthdata.com}.
\end{acknowledgments}


\end{article}
\end{document}